\documentclass[12pt]{article}
\usepackage{amsmath,amssymb}
\usepackage{bm}
\usepackage{graphicx}

\makeatletter
  
  \@addtoreset{equation}{section}
\makeatother

\begin{document}
\baselineskip 17pt 
\allowdisplaybreaks

\thispagestyle{empty}
\vspace*{3.0cm}
\begin{center}
\Large {\bf
Semi-classical correlator for 1/4 BPS Wilson loop \\
and chiral primary operator with large R-charge}

\vspace{0.7cm}

\normalsize
 \vspace{0.4cm}

Takayuki {\sc Enari}\footnote{e-mail address:\ \ 
{\tt cstk11001@g.nihon-u.ac.jp}}
and 
Akitsugu {\sc Miwa}\footnote{e-mail address:\ \ 
{\tt akitsugu@phys.cst.nihon-u.ac.jp}}

\vspace{0.7cm}

{\it 
Department of Physics, College of Science and Technology, 
Nihon University, 1-8-14, \\
Kanda-Surugadai, Chiyoda-ku, Tokyo 101-8308, Japan} \\

\vspace{1cm}
{\bf Abstract} 
\end{center}

We study a holographic description for 
correlation function of 1/4 BPS Wilson loop operator
and 1/2 BPS local operator carrying a large R-charge
of order $\sqrt \lambda$.
We construct a rotating string solution which is 
extended in S$^5$ as well as in AdS$_5$. 
The string solution preserves the 1/8 of the 
supersymmetry as expected from the gauge theory computation.
By evaluating the string action including boundary terms
we show that the string solution reproduces correlation 
function in large $J \sim {\cal O}(\sqrt \lambda)$ limit. 
In addition, we found the second solution for which the 
``size'' of the string becomes larger than the radius of S$^5$.
In the case $J=0$, this solution reduces to 
the previously known unstable string configuration.
The gauge theory side also contains a saddle point 
which is not on the steepest descent path.
We show that the saddle point value matches for 
this case as well.

\vspace{2em}

\newpage
\setcounter{page}{1}

\tableofcontents

\section{Introduction}
\label{sec:Int}

Since the AdS/CFT correspondence was proposed 
\cite{Maldacena:1997re}\cite{Gubser:1998bc}\cite{Witten:1998qj}, 
a wide range of its aspects has been investigated. 
Among these, the correspondence between a Wilson loop and 
semi-classical string propagation in the bulk 
\cite{Rey:1998ik}\cite{Maldacena:1998im}\cite{Drukker:1999zq} 
is an example in which precise agreement 
can be studied explicitly.
In that context, great successes have been made 
first by summing up all the ladder diagrams 
\cite{Erickson:2000af}\cite{Drukker:2000rr}\cite{Semenoff:2001xp}
and then by using the localization technique
\cite{Pestun:2007rz}, which made it possible to compute 
the expectation value of the circular Wilson loop 
for finite $N$ and finite $\lambda$ in the gauge theory side. 

In the gravity side, there have been several attempts to 
go beyond the standard large $N$ and large $\lambda$ limit 
by considering the system in which additional
parameters come into the analysis. 
In \cite{Drukker:2005kx} the number of the string charge is taken to 
be large, which makes it possible to discuss the all genus 
contribution, although the large $N$ limit is assumed.
In \cite{Zarembo:2002ph},  a large angular momentum
\cite{Berenstein:2002jq} is introduced for the study of 
correlation function between the Wilson loop and 
a local operator \cite{Berenstein:1998ij}.
It was found that the semi-classical string analysis
reproduces the gauge theory result which is derived by
summing up all the ladder diagrams \cite{Semenoff:2001xp}.
For semi-classical analysis of the correlation functions
which include Wilson loop and large R-charge see, 
for example, the papers
\cite{Pestun:2002mr}\cite{Alday:2011pf}.
On the other hand, the 1/4 BPS Wilson loop, 
which depends on the parameter corresponding 
to the S$^5$ angle in the gravity side is studied
in \cite{Drukker:2006ga}\cite{Drukker:2006zk}.

In the present paper, we consider the correlation
function between the 1/4 BPS Wilson loop and 
the 1/2 BPS local operator carrying a large R-charge. 
This amounts to introducing both of the parameters 
introduced in \cite{Zarembo:2002ph} and \cite{Drukker:2006ga} 
at the same time.
For this purpose, we need to construct a string solution
which is extended both in the AdS$_5$ and the S$^5$ part
and rotating in the S$^5$. 

Regarding the 1/4 BPS Wilson loop, one of the 
interesting features found in \cite{Drukker:2006ga} is that 
there exist two solutions representing a stable
and an unstable string configuration. 
These solutions reproduce the contributions
coming from two saddle points in the gauge theory side.
One of the motivations for the present work 
is to investigate the same issue in the case 
with large angular momentum.

The paper is organized as follows.
In section \ref{sec:Gauge}, we review the 
results in the gauge theory side and the 
saddle points for the Bessel function.
We study the semi-classical description
in the gravity side in section 
\ref{sec:Semi}.
We start with constructing the string 
solution in subsection \ref{sec:S-g} by using the
global coordinate for the bulk geometry, 
and study its supersymmetry property 
of the solution in subsection \ref{sec:BPS}.
Then in subsections \ref{sec:S-P}--\ref{sec:Eva}, 
we evaluate the string action 
including appropriate boundary terms
and compare it with the gauge theory results.
In subsection \ref{sec:Gen}, we construct a string solutions
corresponding to generic configurations for the Wilson loop
and local operator. 
The correspondence for the solution whose ``size''
becomes larger than the S$^5$, which is found in subsection 
\ref{sec:S-g}, is further investigated in subsection 
\ref{sec:Ano}.
Section \ref{sec:S-D} is devoted to the 
summary and discussion.

\section{Correlation Function in Gauge Theory}
\label{sec:Gauge}

\subsection{Wilson Loop and Local Operator}
\label{sec:W-L}
The main goal of the present paper is to 
study the AdS/CFT correspondence for 
the correlation function between
a 1/4 BPS Wilson loop and a 1/2 BPS local 
operator with large R-charge $J$ of order 
$\sqrt \lambda$\, ($\lambda$ is the 't Hooft coupling), 
\begin{equation}
\langle W(C) {\cal O}_J (\vec x_0) \rangle\,.
\end{equation}
The explicit forms of $W(C)$ and 
${\cal O}_J$ are as follows
\begin{align}
W(C) 
& = {\rm trPexp}
\int_0^{2 \pi} d \sigma
\Big(
i A_\mu \dot x^\mu
+
|\dot x|
\Phi_i \Theta_i
\Big)\,, \\
{\cal O}_J 
& = {\rm tr}\, (\Phi_3 - i \Phi_4)^J\,.
\end{align}
Here the loop $C$ for the Wilson loop is taken to be
a circle on the ($x_1$, $x_2$) plane
\begin{equation}
\vec x(\sigma) = (r \cos \sigma\,,\, r \sin \sigma\,,\, 0\, ,\, 0)\,,
\end{equation}
and the $\sigma$ dependent 6 dimensional 
unit vector $\vec \Theta$ is taken to be
\begin{equation}
\vec \Theta (\sigma) =
(
\sin \theta_0 \cos \sigma\,, \,
\sin \theta_0 \sin \sigma\,, \,
\cos \theta_0 \cos \chi_0\,, \,
\cos \theta_0 \sin \chi_0\,, \,
0\,, 0
)\,,
\end{equation}
where $\theta_0$ and $\chi_0$ are constants. 
The position $\vec x_0$ of the local operator 
is taken to be $\vec x_0 = (0,0,0,\ell)$.
We will study other cases in subsection \ref{sec:Gen} 
from the gravity side, where the Wilson loop is also changed. 
The parameter $\chi_0$ can be 
eliminated from the Wilson loop by the field redefinition
\begin{equation}
\Phi_3' - i \Phi_4'
=
{\rm e}^{i \chi_0}
(\Phi_3 - i \Phi_4)\,.
\end{equation}
After the redefinition, the local operator 
is changed by a phase factor 
\begin{equation}
{\cal O}_J = 
{\rm e}^{-iJ\chi_0} 
{\rm tr}[(\Phi_3' - i \Phi_4')^J]\,,
\end{equation}
and the correlation functions with $\chi_0 \neq 0$
and $\chi_0 = 0$ are related by the factor
\begin{equation}
\langle 
W(C_{\chi_0 \neq 0}) 
{\cal O}_J
\rangle 
=
{\rm e}^{-i J \chi_0}
\langle 
W(C_{\chi_0=0}) 
{\cal O}_J
\rangle\,. \label{W=eW}
\end{equation}
Hence the system is same as the one studied in
\cite{Semenoff:2006am}, except that we consider 
the large R-charge.
The large $N$ limit of the correlation function is computed 
by summing up all the planar ladder diagrams 
and the result is given by the following form 
\cite{Semenoff:2001xp}\cite{Semenoff:2006am}:
\begin{equation}
{
\langle W(C) {\cal O}_J( \vec x_0 ) \rangle 
\over 
\langle W(C) \rangle}
\propto
{r^J \over (r^2 + \ell^2)^J}
\sqrt{ \lambda' J}
{I_J(\sqrt{\lambda'}) \over I_1(\sqrt{\lambda'})}\,,
\label{WW=I}
\end{equation}
where $I_J$ and $I_1$ represent the modified Bessel functions
and  $\lambda' = \lambda \cos^2 \theta_0$. 
In \cite{Semenoff:2006am}, the large $\lambda$ limit of the 
correlation function is reproduced by considering the bulk 
local fields propagating from the AdS boundary 
to the string worldsheet.
In the case with large R-charge, the effect of the 
vertex operator inserted on the worldsheet is 
not negligible and the saddle point for string path integral is changed. 
A resulting solution in the case with 1/2 BPS Wilson loop ($\theta_0=0$)
is derived in \cite{Zarembo:2002ph} and it is shown that the 
large $J (\sim \sqrt{\lambda})$ behavior of the gauge theory 
correlation function is reproduced from 
the semi-classical string propagation.

The semi-classical analysis of the 1/4 BPS system 
shows interesting features such as the existence of 
an unstable saddle point \cite{Drukker:2006ga}.
Since the large R-charge changes the 
string saddle point \cite{Zarembo:2002ph}, it is interesting to 
study the structure of the saddle point in this case
both in the gauge theory and the string theory.

\subsection{Saddle Points for Bessel Function}
\label{sec:Sad}
Before studying saddle points in the string theory, 
we consider the structure of the saddle points
for the Bessel function \cite{Watson:1944}.
Let us take the following integral representation 
of the modified Bessel function 
\begin{equation}
I_J \big( \sqrt {\lambda'} \big) = 
{1 \over 2 \pi i}
\int_{\infty - \pi i}^{\infty + \pi i}
{\rm e}^{ \sqrt{\lambda'}\cosh z - J z} 
dz\,. \label{IJ=int}
\end{equation}
Here the original contour is defined by 
$z = u - i \pi$ ($ -\infty < u < 0$),
$z = i u $ ($ -\pi < u < \pi$)
and $z = u + i \pi$ ($ 0< u < \infty$). 
The saddle points are located at the roots 
of the equation
\begin{equation}
\sinh z - j' = 0\,, \qquad 
\Big( j' = {J \over \sqrt{\lambda'}}\Big)\,.
\end{equation}
The solutions for the saddle point equation are given by 
\begin{equation}
z = 
\begin{cases}
\phantom{-}\xi' + 2n\pi i\,, \\
-\xi' + (2n + 1) \pi i \,,
\end{cases}
\quad (n=0, \pm 1, \pm 2, \cdots) \,,
\end{equation}
where $\xi' = \log\big(\sqrt{j'^2+1} + j'\big) $.
We consider the range $ -\pi < {\rm Im}[z] \leq \pi$.
We can choose the steepest descent path which comes from 
$\infty - \pi i$ and passes through the saddle point 
$z=\xi'$ then goes to $\infty + \pi i$. 
The leading behavior of the Bessel function
evaluated at the saddle point is given by 
\begin{equation}
I_J \big(\sqrt{\lambda'}\big) \sim {\rm e}^{\sqrt{\lambda'}
\big(\sqrt{{j'}^2+1} + j' \log( \sqrt{j'^2 + 1} - j') \big)}\,.
\label{IJ}
\end{equation}
This is a generalization of the analysis of 
\cite{Zarembo:2002ph} to include 
the parameter $\theta_0$, 
namely the parameters are changed
as $\lambda' = \lambda \cos^2 \theta_0$
and $j' = j/\cos \theta_0$.

Since the remaining saddle points are not on the 
steepest descent path we take, 
they do not contribute to the asymptotic
form of the modified Bessel function.
However as we change the parameter $j' \to 0$, 
the steepest descent path is deformed to
go through the points $z = \pm \pi i$,
and at the same time the saddle points 
$z = - \xi' \pm \pi i$ are shifted to the points.
Indeed this limit corresponds to the case studied 
in \cite{Drukker:2006ga}, where it was found that
there are two string solutions in the gravity side
and that they reproduce the contribution 
from the two saddle points of the 
modified Bessel function $I_1(\sqrt{\lambda'})$\,.
Now we understand that the structure of the saddle points
are changed by the effect of the large parameter $J$. 
So, it is interesting to ask what happens for 
the unstable solution discussed in \cite{Drukker:2006ga}, 
since now the corresponding saddle point
in the gauge theory side is not 
on the steepest descent path for the integral \eqref{IJ=int}. 

Before ending this section, we give the expression 
of the saddle point value for the second saddle point:
\begin{equation}
{\rm e}^{\sqrt{\lambda'} \cosh z - J z} \bigg|_{z  = - \xi' + \pi i}
=
(-1)^J
{\rm e}^{\sqrt{\lambda'}
\big( - \sqrt{{j'}^2+1} - j' \log( \sqrt{j'^2 + 1} - j') \big)}\,.
\end{equation}

\section{Semi-classical Computation in Gravity Side}
\label{sec:Semi}

\subsection{Solution in Global Coordinate}
\label{sec:S-g}
In the gravity side, we start with the following coordinate 
system for the Lorentzian AdS$_5 \times$S$^5$:
\begin{align}
ds^2 & =
L^2 
\Big\{
-\cosh^2 \!\rho dt^2 + d \rho^2 + 
\sinh^2 \!\rho
\big(
d\varphi_1^2 
+ \sin^2 \varphi_1 d\varphi_2^2 
+ \cos^2 \varphi_1 d \varphi_3^2
\big)
\notag \\
& \hspace{1.5cm}
+
d \theta^2 
+ 
\sin^2 \theta d\phi^2
+
\cos^2 \theta 
\big(
d \chi_1^2
+
\sin^2 \chi_1 d\chi_2^2
+
\cos^2 \chi_1 d \chi_3^2
\big)
\Big\}\,.
\label{AdS-S-metric}
\end{align}
We suppose that the Wilson loop on the boundary $\rho=\infty$ is
located at $t=0$, while the local operator is at $t = \infty$ 
and $\rho=0$.
Then the string propagates between them.
For the S$^5$ part it is useful to consider 
the embedding coordinates
\begin{equation}
{\cal X}_1 + i {\cal X}_2 = \sin \theta {\rm e}^{i \phi}\,, \quad
{\cal X}_3 + i {\cal X}_4 = \cos \theta \sin \chi_1 {\rm e}^{i \chi_2}\,, \quad
{\cal X}_5 + i {\cal X}_6 = \cos \theta \cos \chi_1 {\rm e}^{i \chi_3}\,.
\end{equation}
For a fixed worldsheet time, $\tau$, the string is a circle
on the (${\cal X}_1$,\,${\cal X}_2$) plane and localized on other planes.
As for the $\tau$ dependence, the string is rotating on 
(${\cal X}_3$,\,${\cal X}_4$) plane and also the ``size'' $\sin \theta$ of 
the string changes with respect to $\tau$. Namely the 
initial size is $\sin \theta_0$, which is specified by 
the Wilson loop, while in the late time
the string shrinks to a point $\sin \theta \to 0$ $(\tau \to \infty)$ 
and rotates along the great circle on the (${\cal X}_3$,\,${\cal X}_4$)
plane.
Based on these assumptions, we take the following ansatz:
\begin{align}
& 
t = t(\tau)\,, \quad
\rho = \rho(\tau)\,, \quad 
\varphi_1 = {\pi \over 2}\,, \quad 
\varphi_2 = \sigma\,, 
\label{AnsatzAdS}
\\
& 
\theta = \theta(\tau)\,, \quad
\phi = \sigma\,, \quad
\chi_1 = {\pi \over 2}\,, \quad 
\chi_2 = \chi_2(\tau)\,.
\label{AnsatzS}
\end{align} 
This ansatz corresponds to the 
AdS$_3 \times$S$^3$ ansatz in \cite{Drukker:2005cu}.

The Polyakov action in the conformal gauge is given by 
\begin{equation}
S = {\sqrt \lambda \over 2 }\int d\tau 
\bigg\{
- \cosh^2 \rho (\partial_\tau t)^2 
+
(\partial_\tau \rho)^2
-
\sinh^2 \rho
+
\cos^2 \theta (\partial_\tau \chi_2)^2
+
(\partial_\tau \theta)^2
-
\sin^2 \theta
\bigg\}\,.
\end{equation}
The equations of motion are
\begin{align}
& \partial_\tau 
\big(
\cosh^2 \rho 
\partial_\tau t
\big)
= 0\,, \label{EOMtt}\\
& 
\partial_\tau^2 \rho 
+
\sinh \rho \cosh \rho 
\big( (\partial_\tau t)^2 + 1 \big)
=
0\,, \label{EOMr}\\
& 
\partial_\tau 
\big(
\cos^2 \theta \partial_\tau \chi_2
\big)
= 0\,, \label{EOMc}\\ 
& 
\partial_\tau^2 \theta
+
\sin \theta \cos \theta 
\big(
(\partial_\tau \chi_2)^2 + 1
\big) 
= 0\,. \label{EOMt}
\end{align}
From these equations, we obtain four constants of motion:
\begin{align}
& C_1 = \cosh^2 \rho \partial_\tau t\,,
\label{C_1}\\
&C_2 = \cos^2 \theta \partial_\tau \chi_2\,,
\label{C_2}\\
& 
C_3 = 
- \cosh^2 \rho (\partial_\tau t)^2 
+ (\partial_\tau \rho)^2
+ \sinh^2 \rho\,,
\label{C_3}\\
& 
C_4=\cos^2 \theta (\partial_\tau \chi_2)^2
+ (\partial_\tau \theta)^2
+ \sin^2 \theta\,.
\label{C_4}
\end{align}
The Virasoro-Hamiltonian constraint imposes
the condition $C_3 + C_4 =0$\,.
Since the each constant $C_1$ and $C_2$ 
corresponds to the conformal weight and 
the R-charge of the operator ${\cal O}_J$, respectively, 
we have $C_1 = C_2 = J/\sqrt \lambda \equiv j$.
Then our assumption $(\sin \theta, \rho) \to (0,0)$ 
in the limit $\tau \to \infty$
consistently fixes all the constants:
\begin{equation}
C_1 = j\,, \quad 
C_2 = j\,, \quad
C_3 = - j^2\,, \quad
C_4 = j^2\,.
\end{equation}
By substituting these and \eqref{C_1}, \eqref{C_2}
into \eqref{C_3} and \eqref{C_4},
we obtain equations for $\rho$ and $\theta$
\begin{align}
& (\partial_\tau \rho)^2 
= 
- \sinh^2 \rho - j^2 \tanh^2 \rho\,, \\
& (\partial_\tau \theta)^2
=
- \sin^2 \theta - j^2 \tan^2 \theta\,,
\end{align}
which clearly shows that there is 
no real solution satisfying the boundary condition.
Therefore, we consider Wick rotation 
$\tau_{\rm E} = i \tau$ and $t_{\rm E}=i t$
and look for the Euclidean solution 
\begin{align}
& (\partial_{\tau_{\rm E}} \rho)^2 
= 
\sinh^2 \rho + j^2 \tanh^2 \rho\,, \label{EOMrE}\\
& (\partial_{\tau_{\rm E}} \theta)^2
=
\sin^2 \theta + j^2 \tan^2 \theta\,. \label{EOMtE}
\end{align}
Now it is not difficult to find 
the following solutions for these equations
\begin{align}
&
\sinh \rho 
= {\sqrt{j^2 + 1} \over \sinh\sqrt{j^2+1} \tau_{\rm E}}\,, 
\label{SOL-rho}\\
&
\sin \theta 
= {\sqrt{j^2 + 1} \over \cosh\sqrt{j^2+1}
(\tau_{\rm E} + \tau_0)}\,. \label{SOL-theta}
\end{align}
Here the constant $\tau_0$ ($\geq 0 $) is related to the 
parameter $\theta_0$ in the gauge theory side by 
\begin{equation}
\sin \theta_0 = {\sqrt{j^2+1} \over \cosh \sqrt{j^2+1} \tau_0}\,.
\end{equation}
The equations for $t_{\rm E}$ and $\chi_2$ can 
be solved and the solutions are given by
\begin{align}
&
t_{\rm E} = 
j \tau_{\rm E} 
- 
{1 \over 2} 
\log\bigg(
{
\cosh(\sqrt{j^2+1}\tau_{\rm E}+\xi) 
\over 
\cosh(\sqrt{j^2+1}\tau_{\rm E}-\xi)} 
\bigg)\,,
\label{SOL-t}\\
& 
\chi_2 = 
- i j \tau_{\rm E}
+
{i \over 2}
\log
\bigg(
{
\sinh(\sqrt{j^2+1}(\tau_{\rm E}+\tau_0) + \xi)
\sinh(\sqrt{j^2+1} \tau_0 - \xi)
\over
\sinh(\sqrt{j^2+1}(\tau_{\rm E}+\tau_0) - \xi)
\sinh(\sqrt{j^2+1} \tau_0 + \xi)
}
\bigg)
+
\chi_2(0)\,, \label{SOL-chi2}
\end{align}
where $\xi = \log(\sqrt{j^2+1} + j)$. 
The initial value of $\chi_2$ is set to be
$\chi_2(0) = \chi_0$, while that for $t_{\rm E}$ is taken to be zero,
$t_{\rm E}(0) = 0$. 
Here since the angular momentum $J$ (and hence $j$)
is kept to be real, the solution for $\chi_2$ becomes imaginary
after Wick rotation. It is usual for the semi-classical analysis
with large angular momentum \cite{Dobashi:2002ar}\cite{Zarembo:2002ph}.

From \eqref{EOMtE}, we see that there is yet another solution 
\begin{equation}
\sin \theta = 
{\sqrt{j^2 + 1} \over \cosh \sqrt{j^2 + 1} (-\tau_{\rm E}+\tau_0)}\,, 
\label{another}
\end{equation}
which satisfies the boundary condition; 
$\sin \theta(0) = \sin \theta_0$ and 
in the late time it shrinks to a point, 
i.e., $\sin \theta(\infty) = 0$. 
For this solution, $\sin \theta$ becomes larger than $1$ 
which is out side of the original integral domain. 
However since we are now considering analytic continuation, 
the meaning of the integral domain is not very clear.
In subsection \ref{sec:Ano}, we study the property of the
second solution and find that it corresponds to 
the saddle point of the Bessel function, 
which is not on the steepest descent path 
taken in the saddle point analysis in subsection \ref{sec:Sad}.
Note that in the case $J=0$, it reduces to the unstable string
solution discussed in \cite{Drukker:2006ga}.

\subsection{BPS Condition}
\label{sec:BPS}
The BPS condition in the gauge theory side is
addressed in \cite{Semenoff:2006am}
and it was found that the system is $1/8$ BPS. 
Now we study the supersymmetry preserved by the
solution discussed in the previous section.

The vielbeins are taken as follows:
\begin{equation}
\begin{cases}
e^0 = L \cosh \rho dt\,, \\
e^1 = L d \rho\,, \\
e^2 = L \sinh \rho d \varphi_1\,, \\
e^3 = L \sinh \rho \sin \varphi_1 d \varphi_2\,, \\
e^4 = L \sinh \rho \cos \varphi_1 d \varphi_3\,, \\
\end{cases}
\quad 
\begin{cases}
e^5 = L d \theta\,, \\
e^6 = L \sin \theta d \phi\,, \\
e^7 = L \cos \theta d \chi_1\,, \\
e^8 = L \cos \theta \sin \chi_1 d \chi_2\,, \\
e^9 = L \cos \theta \cos \chi_1 d \chi_3\,.
\end{cases}
\end{equation}
The BPS condition in the Lorentzian signature is given by 
\begin{equation}
{1 \over \sqrt{-\det g}}
\partial_\tau X^M \partial_\sigma X^N  \hat \Gamma_M \hat \Gamma_N 
\sigma_3 \epsilon = \epsilon\,, \label{kappaL}
\end{equation}
where $g$ is induced metric on the worldsheet, 
$\hat \Gamma_M$ is defined by $\hat \Gamma_M = e_M^a \Gamma_a$
with constant 10-dimensional gamma matrices $\Gamma_a$, 
and $\epsilon$ is the Killing spinor on AdS$_5 \times$S$^5$. 
We use the notation in which the two Majorana-Weyl spinors with 
common chirality, $\epsilon_1$, $\epsilon_2$
are combined into a column vector $\epsilon$ 
and the two by two matrix $\sigma_3$ acts on it, 
namely, in our notation, 
\begin{equation}
\epsilon = 
\left(
\begin{array}{c}
\epsilon_1 \\
\epsilon_2
\end{array}
\right)\,, \qquad
\sigma_3 \epsilon = 
\left(
\begin{array}{c}
\epsilon_1 \\
- \epsilon_2
\end{array}
\right)\,.
\end{equation}
Another notation often used in literature is to combine two Majorana-Weyl 
spinors into a single complex spinor and use the 
complex conjugation $K \epsilon = \epsilon^\ast$ 
instead of the matrix notation $\sigma_3$.
Since Wick rotation necessarily introduces $i$, 
which is not related to the operation in the space 
of two spinors, $\epsilon_1$ and $\epsilon_2$, 
we use the above notation in order to avoid 
possible confusion.

For $\epsilon$ we take the following form:
\begin{equation}
\epsilon = 
{\rm e}^{{\rho \over 2} \varepsilon \Gamma_\star \Gamma_1}
{\rm e}^{{t \over 2} \varepsilon \Gamma_\star \Gamma_0} 
{\rm e}^{{\varphi_2 \over 2} \Gamma_{13}} 
{\rm e}^{{\theta \over 2} \varepsilon \Gamma_\star \Gamma_5}
{\rm e}^{{\phi \over 2} \Gamma_{56}}
{\rm e}^{{\chi_2 \over 2} \varepsilon \Gamma_\star \Gamma_8}
\Big(
\begin{array}{c}
\epsilon_1^{0} \\
\epsilon_2^{0} 
\end{array}
\Big)\,. \label{Killing}
\end{equation}
Here $\Gamma_\star = \Gamma^{01234}$,  
while $\epsilon_1^0$ and $\epsilon_2^0$ are 
constant Majorana-Weyl spinors with common chirality.
The two by two matrix $\varepsilon = - i \sigma_2$ 
acts on a generic spinor ${}^{\rm t}(\psi_1, \psi_2)$ as 
$\varepsilon\,{}^{\rm t}(\psi_1, \psi_2) = {}^{\rm t}(\psi_2, -\psi_1)$. 
The spinor $\epsilon$ in \eqref{Killing} is the relevant part of the 
Killing spinor satisfying the following equation 
\begin{equation}
\bigg(D_M  - {\varepsilon \over 2 L} \Gamma_\star \hat \Gamma_M \bigg)
\epsilon = 0\,.
\end{equation}

By Wick rotation 
$\tau_{\rm E} = i \tau$, $t_{\rm E} = i t$, 
the BPS condition \eqref{kappaL} becomes 
\begin{equation}
\Gamma \epsilon =
{i \over \sqrt{\det g}}
\partial_{\tau_E} X^M 
\partial_\sigma X^N
\hat \Gamma_M \hat \Gamma_N \sigma_3 \epsilon 
=
\epsilon\,.
\label{project2}
\end{equation}
After Wick rotation, the spinors $\epsilon_1^0$ and 
$\epsilon_2^0$ are not assumed to be Majorana but we
regard them as complex spinors.
This does not mean that we have doubled the degrees of 
freedom, 
because the complex conjugate of the spinor, $\epsilon^\ast$,
will not appear in the following discussion.
We regard this procedure as an analytic continuation 
for the spinor just like other bosonic string coordinates.

By using the ansatz \eqref{AnsatzAdS}, \eqref{AnsatzS}
and Virasoro constraints, we obtain 
\begin{equation}
\Gamma =
{i \over \sinh^2 \rho + \sin^2 \theta }
(
\dot t_{\rm E} \cosh \rho \Gamma_{\rm E}
+
\dot \rho \Gamma_1
+
\dot \theta \Gamma_5
+
\dot \chi_2 \cos \theta \Gamma_8
)
(\sinh \rho \Gamma_3 + \sin \theta \Gamma_6)
\sigma_3\,.
\label{project3}
\end{equation}
Here dots represent the derivative with respect to 
the Euclidean time $\tau_{\rm E}$.
We first eliminate the $\sigma$ dependence of
the Killing spinor \eqref{Killing} by imposing the condition 
\begin{equation}
(\Gamma_{13} + \Gamma_{56}) 
\Big(
\begin{array}{c}
\epsilon_1^0 \\
\epsilon_2^0
\end{array}
\Big)
=
0\,.
\label{condition1}
\end{equation}
Then $\epsilon$ becomes
\begin{equation}
\epsilon = 
{\rm e}^{{\rho \over 2} \varepsilon \Gamma_\star \Gamma_1}
{\rm e}^{{t_{\rm E} \over 2} \varepsilon \Gamma_\star \Gamma_{\rm E}} 
{\rm e}^{{\theta \over 2} \varepsilon \Gamma_\star \Gamma_5}
{\rm e}^{{\chi_2 \over 2} \varepsilon \Gamma_\star \Gamma_8}
\Big(
\begin{array}{c}
\epsilon_1^0 \\
\epsilon_2^0 
\end{array}
\Big)\,.
\end{equation}
Next we multiply the factor 
$
{\rm e}^{- {\rho \over 2} \varepsilon \Gamma_\star \Gamma_1}
{\rm e}^{- {\theta \over 2} \varepsilon \Gamma_\star \Gamma_5}
$
on each side of \eqref{project2}. 
Then it becomes 
$\bar \Gamma \bar \epsilon = \bar \epsilon $ with
\begin{align}
& \bar \Gamma = 
{i \over \sinh^2 \rho + \sin^2 \theta}
\Big[
\dot t_{\rm E} \cosh \rho \sin \theta \Gamma_{{\rm E}6}
+
\dot \chi_2 \sinh \rho \cos \theta \Gamma_{83} 
+
{\rm e}^{- \theta \varepsilon \Gamma_\star \Gamma_5}
\sinh \rho
(
\dot \rho \Gamma_{13} 
+
\dot \theta \Gamma_{53}
) \notag \\
& 
+
{\rm e}^{- \rho \varepsilon \Gamma_\star \Gamma_1}\!
\sin \theta
(
\dot \rho \Gamma_{16}
+
\dot \theta \Gamma_{56}
) 
+
{\rm e}^{-\theta \varepsilon \Gamma_\star \Gamma_5
- \rho \varepsilon \Gamma_\star \Gamma_1}
(
\dot t_{\rm E} \cosh \rho \sinh \rho \Gamma_{{\rm E}3}
+
\dot \chi_2 \cos \theta \sin \theta \Gamma_{86}
)
\Big]\sigma_3\,,
\end{align}
and
\begin{equation}
\bar \epsilon = 
{\rm e}^{{t_{\rm E} \over 2} \varepsilon \Gamma_\star \Gamma_{\rm E}}
{\rm e}^{{\chi_2 \over 2} \varepsilon \Gamma_\star \Gamma_8}
\Big( 
\begin{array}{c}
\epsilon_1^0 \\
\epsilon_2^0
\end{array}
\Big)\,.
\end{equation}
After some calculation, $\bar \Gamma$ acting on 
$\bar \epsilon$ is further rewritten as
\begin{align}
& \bar \Gamma 
=
{i \over \sinh^2 \rho + \sin^2 \theta}
\Big[
\Gamma_{13}
{\rm e}^{\rho \varepsilon \Gamma_\star \Gamma_1}
{\rm e}^{-\theta \varepsilon \Gamma_\star \Gamma_5}
( 
\dot t_{\rm E} \cosh^2 \rho \varepsilon \Gamma_\star \Gamma_{\rm E}
+
\dot \chi_2 \cos^2 \theta \varepsilon \Gamma_\star \Gamma_8 
)\notag \\
& + 
\partial_{\tau_{\rm E}}
\big(
\sinh \rho \sin \theta \Gamma_{16}
+
\cosh \rho \cos \theta \Gamma_{13}
\big)
-
\cosh \rho \cos \theta \Gamma_{13}
(
\dot t_{\rm E} \varepsilon \Gamma_\star \Gamma_{\rm E} 
+ 
\dot \chi_2 \varepsilon \Gamma_\star \Gamma_8)
\Big] \sigma_3\,,
\label{bGamma-2}
\end{align}
where we have used the first condition \eqref{condition1}.
By using the equations of motion $\dot t_{\rm E} \cosh^2 \rho = j$
and $\dot \chi_2 \cos^2 \theta = - i j$, 
the round bracket in the first line in \eqref{bGamma-2} becomes
$j (\varepsilon \Gamma_\star \Gamma_{\rm E} - i \varepsilon \Gamma_\star
\Gamma_8)$. 
Now we impose the second condition 
which is natural for the string with angular momentum on the S$^5$:
\begin{equation}
(\Gamma_{\rm E} - i \Gamma_8)
\Big(
\begin{array}{c}
\epsilon_1^0 \\
\epsilon_2^0
\end{array}
\Big) = 0\,. \label{condition2}
\end{equation}
Then by multiplying the factor 
$
{\rm e}^{-{t_{\rm E} \over 2} \varepsilon \Gamma_\star \Gamma_{\rm E}}
{\rm e}^{-{\chi_2 \over 2} \varepsilon \Gamma_\star \Gamma_8}
$, 
the BPS condition is reduced to 
\begin{equation}
{i \over \sinh^2 \rho + \sin^2 \theta}
\partial_{\tau_{\rm E}} 
\Big(
\sinh \rho \sin \theta\Gamma_{16}
+ \cosh \rho \cos \theta \Gamma_{13}
{\rm e}^{-\varepsilon \Gamma_\star 
(t_{\rm E} \Gamma_{\rm E} + \chi_2 \Gamma_8)}\Big)\sigma_3
\Big(
\begin{array}{c}
\epsilon_1^0 \\
\epsilon_2^0
\end{array}
\Big)
= 
\Big(
\begin{array}{c}
\epsilon_1^0 \\
\epsilon_2^0
\end{array}
\Big). \label{project4}
\end{equation}
Since both of the solutions found in subsection \ref{sec:S-g},
i.e., \eqref{SOL-theta} and \eqref{another},
satisfy the relation
\begin{equation}
\partial_{\tau_{\rm E}}
( \sinh \rho \sin \theta )
=
-\sin \theta_0 (\sinh^2 \rho + \sin^2 \theta)\,,
\label{sinh2+sin2}
\end{equation}
the second term in the round bracket 
on the left hand side of \eqref{project4} is 
expected to be proportional 
to $\sinh \rho \sin \theta$ up to some additive constant.
This is indeed the case and we have
\begin{equation}
\cosh \rho \cos \theta 
{\rm e}^{-(t_{\rm E} - i \chi_2) \varepsilon \Gamma_\star \Gamma_{\rm
E}}
=
\bigg(
\sinh \rho \sin \theta
{\cos \theta_0 \over \sin \theta_0}
\pm
{\sqrt{{j^2 \over \cos^2 \theta_0}+ 1}  }
+
{j \over \cos \theta_0} \varepsilon \Gamma_\star \Gamma_{\rm E}
\bigg)
{\rm e}^{i \chi_2(0) \varepsilon \Gamma_\star \Gamma_{\rm E}}
\,. \label{cohrcostE}
\end{equation}
Here the plus sign is for the first solution for which the S$^5$ 
part is given by \eqref{SOL-theta} and \eqref{SOL-chi2},
while the minus sign is for the second solution \eqref{another},
the explicit form of $\chi_2$ for this solution is given 
in subsection  \ref{sec:Ano}.
The constant terms, the second and the third terms 
in the round bracket of \eqref{cohrcostE}, 
will drop from \eqref{project4}
because of the derivative.
Then the BPS condition is satisfied if 
the following third condition is imposed:
\begin{equation}
- i (
\sin \theta_0 \Gamma_{16} 
+ 
\cos \theta_0 \Gamma_{13}{\rm e}^{  i  \chi_2(0)\varepsilon
\Gamma_\star \Gamma_{\rm E}}
) \sigma_3
\Big(
\begin{array}{c}
\epsilon_1^0 \\
\epsilon_2^0
\end{array}
\Big)
=
\Big(
\begin{array}{c}
\epsilon_1^0 \\
\epsilon_2^0
\end{array}
\Big)\,. \label{condition3}
\end{equation}

In summary, since the projections \eqref{condition1}, 
\eqref{condition2} and \eqref{condition3} commute
with each other, both of the string solutions found in the 
previous subsection preserve 1/8 of the supersymmetry.
The conditions \eqref{condition1}, \eqref{condition3}
are the ones discussed in \cite{Drukker:2006ga}
and another condition \eqref{condition2} is for the
rotating string.

\subsection{Solution in Poincar\'e AdS}
\label{sec:S-P}
The solution in the previous section is 
useful for the analysis of its symmetry property, 
but it is not suited to study the holographic
correspondence to the gauge theory. 
This is because the local operator is located at infinity, 
and also the size of the loop is not clear.
In order to study the correspondence, including 
the dependence on these parameters, we need to 
construct the solution in Poincar\'e coordinate
and put the local operator and the Wilson loop
within a finite distance.

In the Euclidean signature, the global coordinate and 
the Poincar\'e coordinate are related through the 
following simple coordinate redefinition:
\begin{equation}
Y = {r {\rm e}^{t_{\rm E}} \over \cosh \rho}\,, \quad
R = r {\rm e}^{t_{\rm E}} \tanh \rho\,, 
\label{global-Poincare}
\end{equation}
where $r$ is a constant parameter. 
The AdS metric in \eqref{AdS-S-metric} is changed to 
\begin{equation}
ds^2 
=
L^2 {d Y^2 + (d X^i)^2 \over Y^2}
=
L^2
{
d Y^2 + d R^2 + R^2 ( d \varphi_1^2 + \sin^2 \varphi_1 d \varphi_2^2
+ \cos^2 \varphi_1 d \varphi_3^2) \over Y^2
}\,.
\end{equation}
The flat 4-dimensional coordinate $\vec X$ is introduced as 
\begin{equation}
 \vec X =
R ( 
\sin \varphi_1 \cos \varphi_2, \,\,
\sin \varphi_1 \sin \varphi_2, \,\,
\cos \varphi_1 \cos \varphi_3, \,\,
\cos \varphi_1 \sin \varphi_3
)\,.
\end{equation}
The AdS part of the solution, which is given by
\eqref{SOL-rho} and \eqref{SOL-t}, is mapped to the configuration
\begin{align}
& 
Y 
= 
Y(\tau_{\rm E})
\equiv
r {\rm e}^{j \tau_{\rm E}}
\Big[
\sqrt{j^2 + 1} \tanh 
\Big( \sqrt{j^2 + 1} \tau_{\rm E} + \xi \Big)
- j
\Big]\,, \label{SOL-Y-Z}\\
& R = R(\tau_{\rm E}) \equiv 
{r {\rm e}^{j \tau_{\rm E}}{ \sqrt{j^2 + 1} } 
\over \cosh(\sqrt{j^2+1}
 \tau_{\rm E} + \xi)}\,, \label{SOL-R-Z}
\end{align}
which is the form found in \cite{Zarembo:2002ph}.
Here we have introduced the functions $Y(\tau_{\rm E})$
and $R(\tau_{\rm E})$ for later convenience.
Although the solution is now in the Poincar\'e coordinate,
the position of the local operator is still at infinity, $Y=\infty$. 
Then we further transform the solution by using the 
isometry of the Poincar\'e AdS
\begin{equation}
\vec X' =
{ 
\vec X + \vec c (\vec X^2 + Y^2) 
\over
1 + 2 \vec c \cdot \vec X + \vec c^2 (\vec X^2 + Y^2)
}\,, \qquad 
Y' =
{
Y
\over
1 + 2 \vec c \cdot \vec X + \vec c^2 (\vec X^2 + Y^2)
}\,. 
\label{isometry}
\end{equation}
This transformation brings the local operator 
to a point a finite distance from the Wilson loop.

Let us consider the symmetric case, i.e., 
the local operator is on an axis of the Wilson loop.
For this purpose we take $\vec c = (0,0,0, 1/ \ell)$.
After the transformation \eqref{isometry}, 
we further perform the translation into ${X^4}'$ 
direction by $-\ell r^2 /(\ell^2 + r^2)$
and also the scale transformation by the factor
$(\ell^2 + r^2)/ \ell^2$, so that the
center of the Wilson loop comes to the origin and
the radius of the loop becomes $r$.
The position of the local operator after the 
whole transformation is $\vec x = (0,0,0,\ell)$.
The explicit form of the combined 
coordinate transformation is given by
\begin{align}
({X^1}',{X^2}',{X^3}', -{X^4}' )
& = {(\ell^2 + r^2) ( \vec X + \vec x )
\over 
(\vec X + \vec x)^2 + Y^2 }
- \vec x\,, \quad 
Y'  
= 
{(\ell^2 + r^2)Y \over 
(\vec X + \vec x)^2 + Y^2 }\,.
\end{align}
The solution \eqref{SOL-Y-Z}, \eqref{SOL-R-Z} is mapped to
\begin{equation}
\vec X'
=
{(\ell^2 + r^2) \over \ell^2 + R^2 + Y^2}
(R \cos \sigma, R \sin \sigma, 0 , - \ell) 
+
(0 , 0 , 0 , \ell)\,, \quad
Y'= {(\ell^2 + r^2) Y \over \ell^2 + R^2 + Y^2}\,.
\label{SOL-(l,r)}
\end{equation}
The each worldsheet boundary, 
at $\tau_{\rm E} = 0$ and $\tau_{\rm E} = \infty$,  
is mapped to 
the circle $(r \cos \sigma, r \sin \sigma, 0 ,0 )$ 
and the point $(0,0,0,\ell)$
on the AdS boundary $Y' = 0$, respectively.
Figure 1 depicts the string worldsheet for 
$j=r = \ell = 1$.

\begin{figure}[htbp]
\begin{center}
\includegraphics[width=6cm]{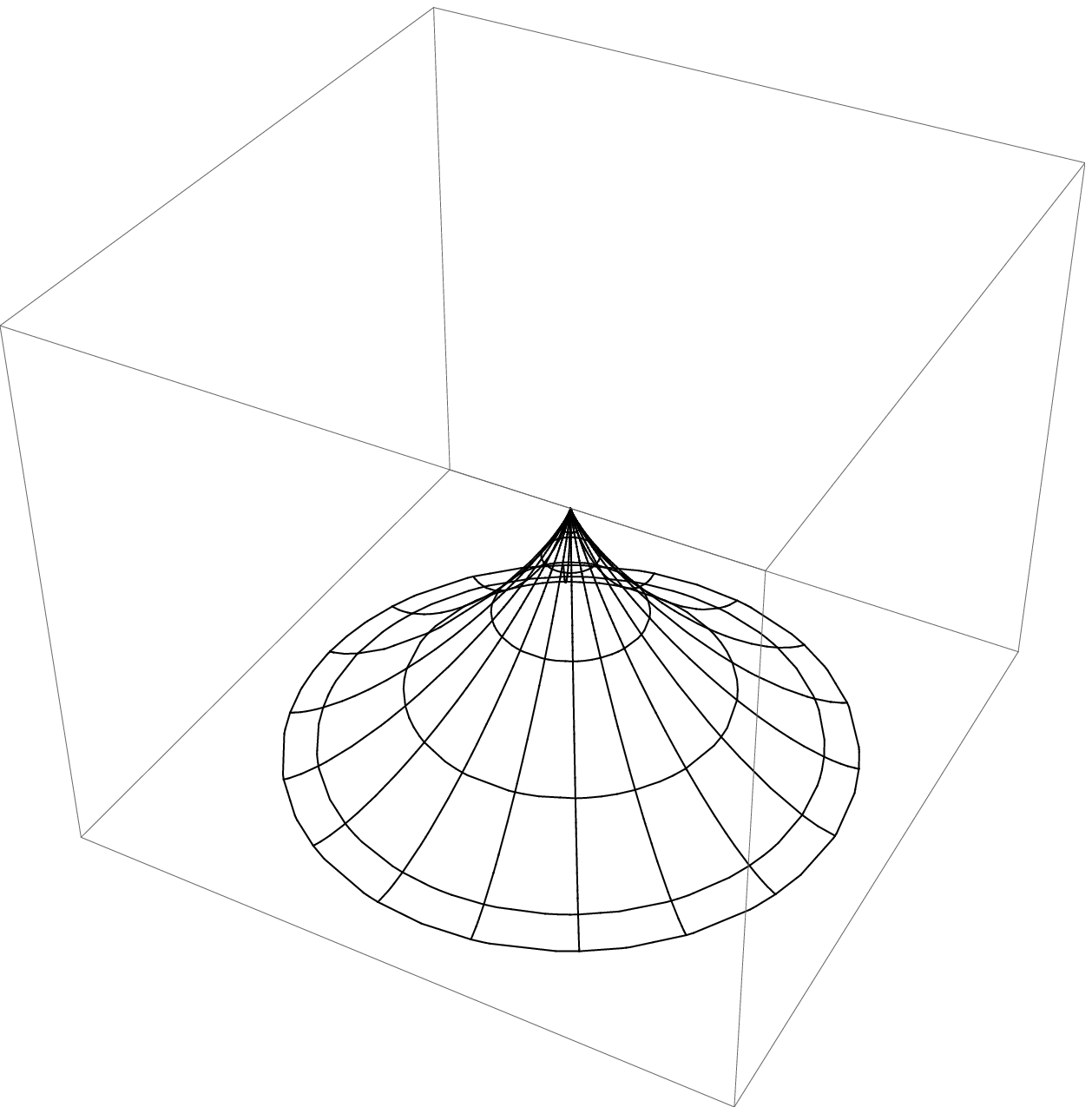} 
\put(-122,8){\footnotesize ${X^1}'$}
\put(-26,33){\footnotesize ${X^2}'$}
\put(-183,75){\footnotesize ${X^4}'$} 
\hspace{1cm}
\includegraphics[width=6cm]{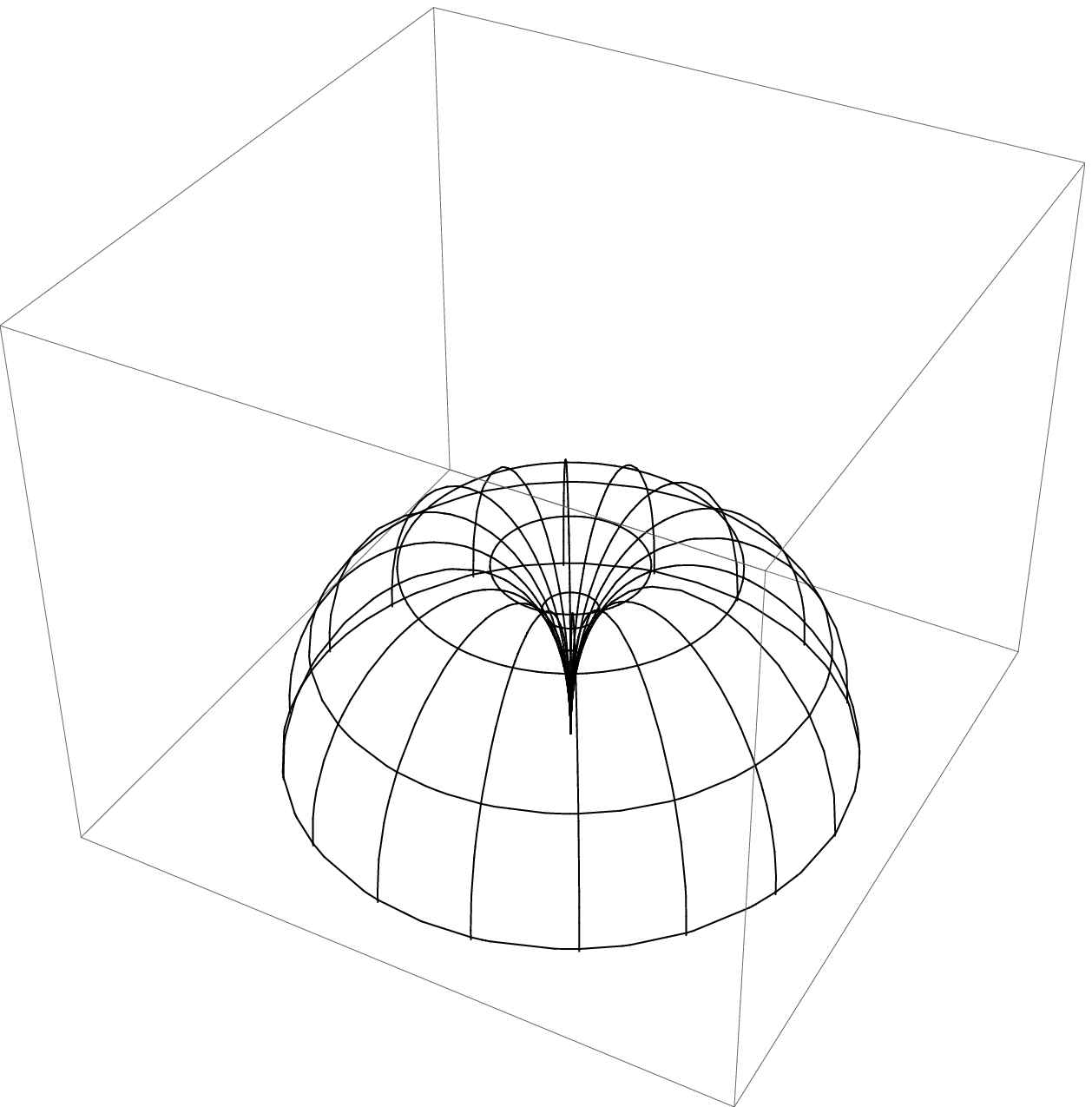}
\put(-122,8){\footnotesize ${X^1}'$}
\put(-26,33){\footnotesize ${X^2}'$}
\put(-181,75){\footnotesize $Y'$}
\caption{The configuration in $({X^1}',{X^2}',{X^4}')$ 
and $({X^1}',{X^2}',Y')$ spaces.
The parameters are $j=r = \ell = 1$.}
\end{center}
\end{figure}

\subsection{Evaluation of Action}
\label{sec:Eva}

In order to study semi-classical string propagation, 
we evaluate the string action with 
boundary terms corresponding to initial 
and final states of the string.
On the Wilson loop side, $\tau_{\rm E} = 0$, 
it is standard to perform the Legendre transformation \cite{Drukker:1999zq}
with respect to the AdS radial coordinate $u=1/Y'$
by adding the following boundary term:
\begin{equation}
S_{{\rm b}, \tau_{\rm E}=0} 
= {\partial L \over \partial \dot u} u \bigg|_{\tau_{\rm E} = 0}.
\label{DGO-boundary}
\end{equation}
Here $L$ is the Lagrangian, i.e., the Euclidean action is given by 
$S_{\rm E} = \int d\tau_{\rm E} L$.
On the other boundary at $\tau_{\rm E} = \infty$, 
we add the boundary term coming from the 
vertex operator $V_J$ corresponding to the local operator
${\cal O}_J$ \cite{Tseytlin:2003ac}\cite{Zarembo:2002ph}%
\cite{Buchbinder:2010vw}:
\begin{equation}
- S_{{\rm b},\tau_{\rm E}=\infty} 
=
\log V_J \Big|_{\tau_{\rm E}=\infty }
=
\bigg[
J \log { Y' \over {{Y'}^2 + (\vec X' - \vec x)^2}}
+
J \log  \cos \theta \sin \chi_1 {\rm e}^{- i \chi_2} 
\bigg]_{\tau_{\rm E}=\infty}\,.
\end{equation}
Here $\sin \chi_1 = 1 $ for the present case.
Since the each boundary term is divergent, we introduce cutoffs
$\tau_{\rm -}$ and $\tau_{\rm +}$ for lower and upper boundaries 
of the integral. 
In summary, we evaluate the following functional
\begin{align}
S_{\rm total} & =
S_{\rm bulk} 
+ S_{{\rm b}, {\tau_{\rm E}=\tau_-} }
+ S_{{\rm b}, {\tau_{\rm E}=\tau_+}}\,. 
\end{align}

For the bulk part $S_{\rm bulk}$,
it is enough to evaluate it for the original solution 
\eqref{SOL-rho}, \eqref{SOL-theta}, \eqref{SOL-t} and \eqref{SOL-chi2}.
\begin{align}
S_{\rm bulk} 
& =
{\sqrt \lambda}
\int_{\tau_-}^\infty d\tau_{\rm E}
( \sinh^2 \rho + \sin^2 \theta ) \\
& =
\sqrt{\lambda}
\Big[
-{1 \over \sin \theta_0} \sinh \rho \sin \theta
\Big]_{\tau_-}^\infty \\
& = 
\sqrt \lambda
\bigg[
{1 \over \tau_-}
-
\sqrt{j^2 + 1}
\tanh \sqrt{j^2 + 1} \tau_0
+ \cdots
\bigg]\\
& = 
\sqrt \lambda
\bigg[
{1 \over \tau_-}
-
\sqrt{j^2 + \cos^2 \theta_0}
+
\cdots
\bigg]\,. \label{Sbulk}
\end{align}
Here the integration is done by using the relation \eqref{sinh2+sin2}.
This divergence is canceled by the first boundary term:
\begin{equation}
{\partial L \over \partial \dot u} u \bigg|_{\tau_-}
=
-\sqrt \lambda {\dot Y' \over Y'} \bigg|_{\tau_-}
=
- \sqrt \lambda
\bigg[
{\dot Y \over Y}
-
{ \partial_{\tau_{\rm E}}(R^2 + Y^2) \over \ell^2 + R^2 + Y^2}
\bigg]_{\tau_-}
=
- {\sqrt \lambda \over \tau_-} 
+ \cdots \,. \label{b-term}
\end{equation}
Here the second term in the square bracket on the 
third expression is zero in the limit $\tau_- \to 0$.
So we have $\dot Y' / Y' \big|_{\tau_-} = \dot Y / Y \big|_{\tau_-}$
in the limit.

The evaluation of the vertex operator is as follows:
\begin{align}
& J \log
{ Y' \over {Y'}^2 + (\vec X' - \vec x)^2}
\Bigg|_{\tau_+} =
J \log
{ Y (\tau_+) \over {\ell^2 + r^2}} 
=
J\bigg[
\log {r \over \ell^2 + r^2}
+
j \tau_+ 
+ \log(\sqrt{j^2 + 1} - j)
+ \cdots
\bigg]\,, \label{EVAL-Vs1}\\
& J \log \cos \theta 
{\rm e}^{-i \chi_2} \bigg|_{\tau_+}
= 
J 
\bigg[
- j \tau_+ -i \chi_0
- \log(\sqrt{j^2 + 1} - j)
+ \log\bigg(\sqrt{{j^2 \over \cos^2 \theta_0}+1} 
- {j \over \cos \theta_0}\bigg) + \cdots
\bigg]\,,
\label{EVAL-Vs2}
\end{align}
where $\chi_0 = \chi_2 (0)$.
Adding all terms, the divergences cancel and we obtain
\begin{align}
{\rm e}^{-{\rm S}_{\rm total}}
& = 
{\rm e}^{-i J \chi_0}
\bigg({r \over \ell^2 + r^2}\bigg)^J
\exp
\sqrt {\lambda'}
\bigg[
\sqrt{{j'}^2 + 1 }
+
j' \log \Big(\sqrt{ {j'}^2 +1 } 
- j'
\Big)
\bigg]\,, \label{e-S}
\end{align}
where $\lambda'= \lambda \cos^2 \theta_0$, 
$j' = j / \cos \theta_0$\,.
This completely reproduces the large $J$ limit
of the gauge theory side including the 
scaling behavior; see the equations \eqref{W=eW},
\eqref{WW=I} and \eqref{IJ}. 
Note that, in \eqref{WW=I}, 
the factor $ (\langle W(C) \rangle)^{-1}$ 
on the left hand side cancels $(I_1(\sqrt{ \lambda'}))^{-1}$
on the right hand side in the large $\lambda'$ limit 
\cite{Drukker:2006ga}, and \eqref{e-S} reproduces the 
remaining factor in the large $J \sim {\cal O}(\sqrt{\lambda'})$ limit.

\subsection{Generic Configuration}
\label{sec:Gen}
By the transformation \eqref{isometry} the local operator
can be placed at arbitrary point.
Since the original solution \eqref{SOL-Y-Z}, \eqref{SOL-R-Z} 
is invariant under the rotation
both on ($X^1$,\,$X^2$) plane and on ($X^3$,\,$X^4$) plane,
we consider the transformation with $\vec c = (c_1, 0, 0, c_4)$.
\begin{figure}[htbp]
\begin{center}
\begin{minipage}{12cm}
\begin{center}
\includegraphics[width=5cm]{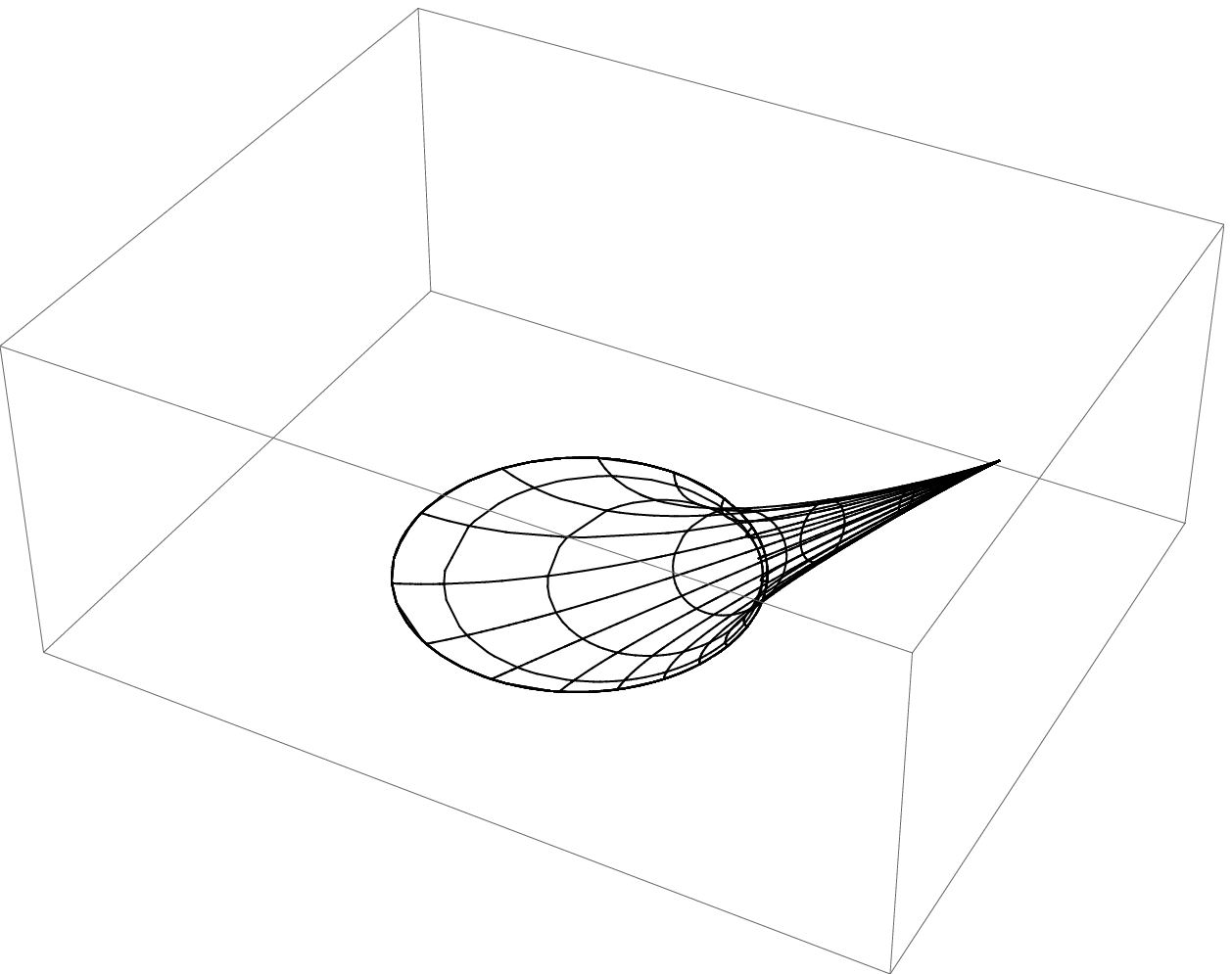} 
\hspace{1cm}
\includegraphics[width=5cm,clip]{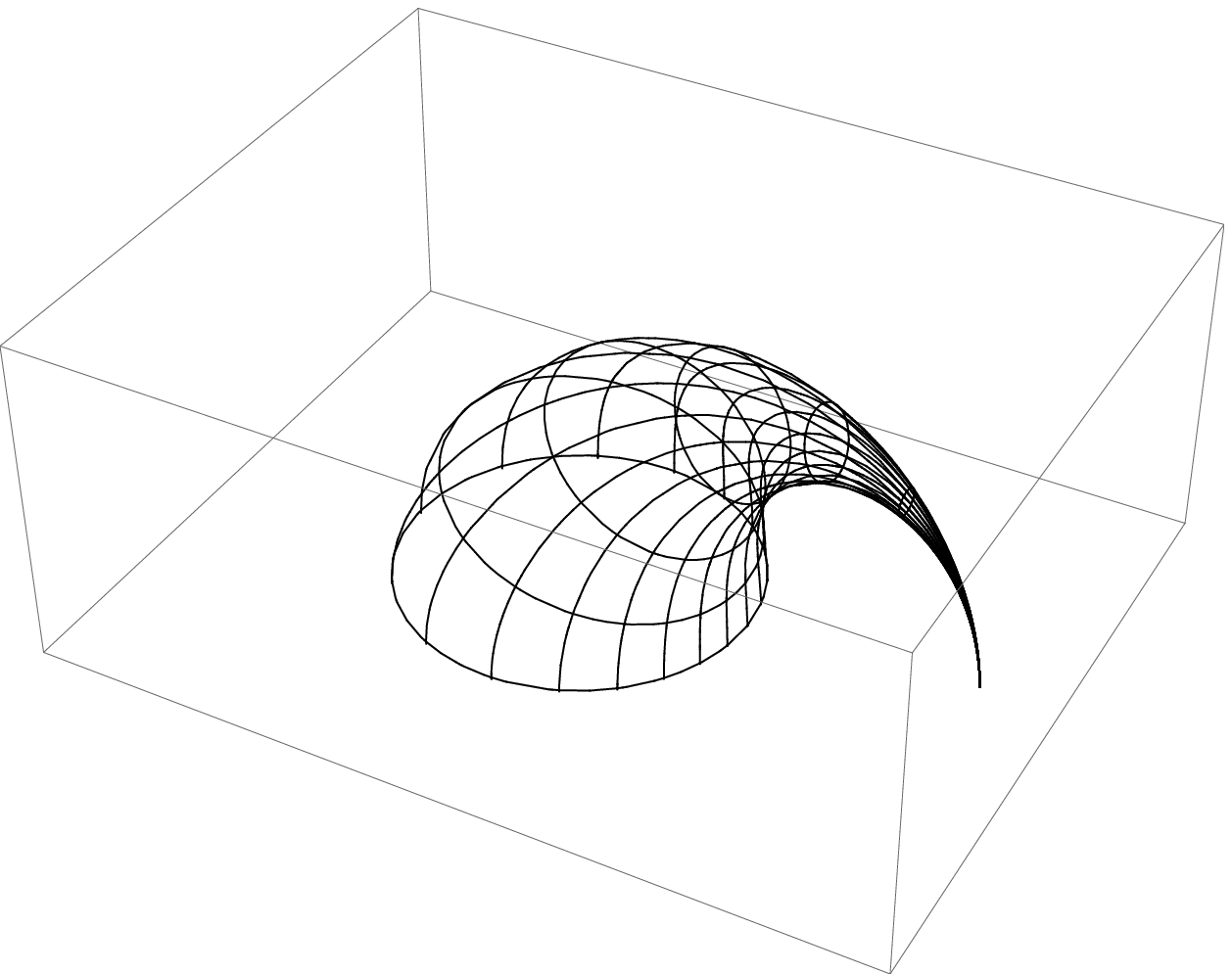}
\caption{The configuration for $(c_1, c_4)=(0.3, 0.2)$
and $j=r=1$.}
\end{center}
\end{minipage} \\[2mm]
\begin{minipage}{12cm}
\begin{center}
\includegraphics[width=5cm,clip]{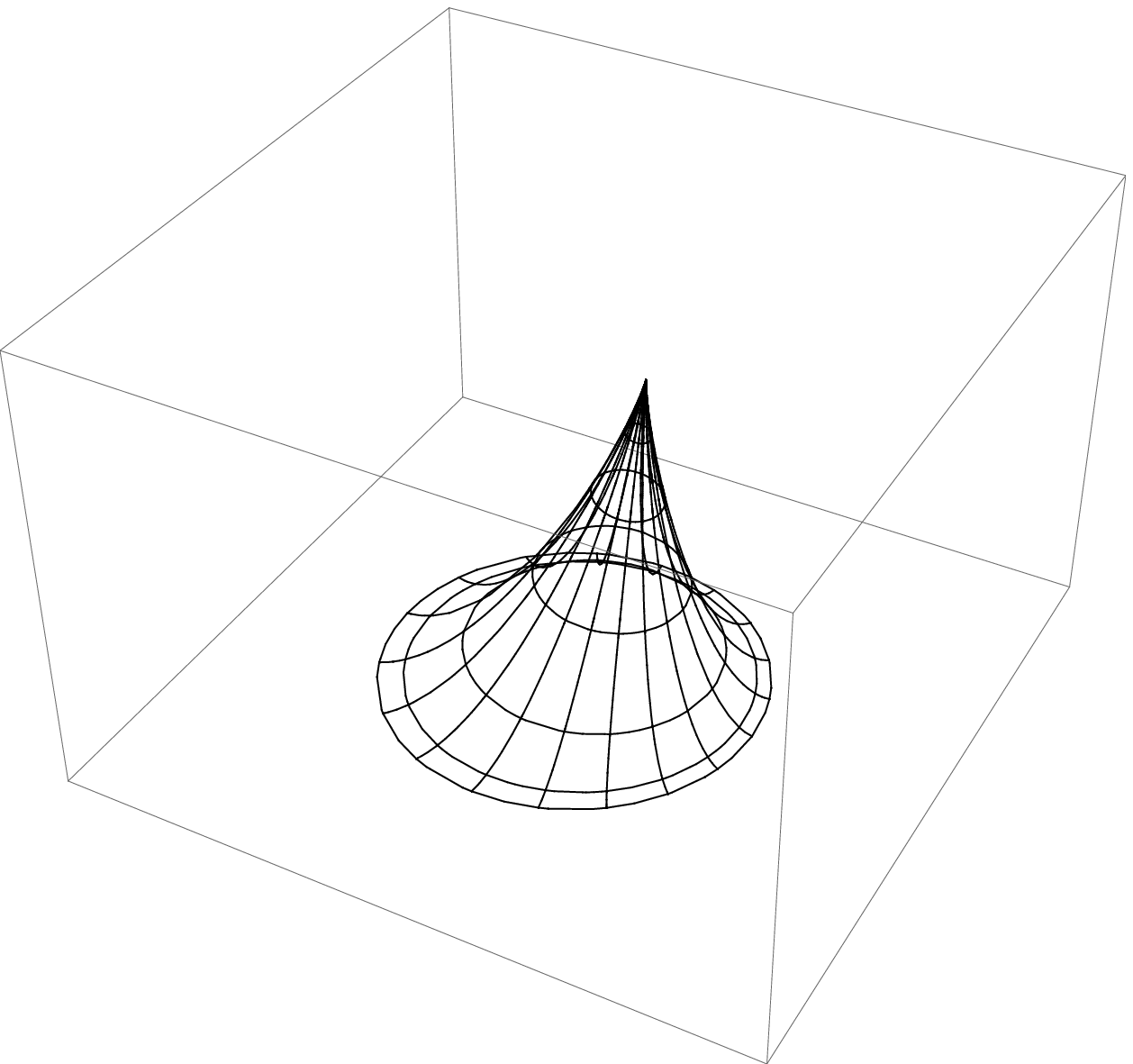}
\hspace{1cm}
\includegraphics[width=5cm,clip]{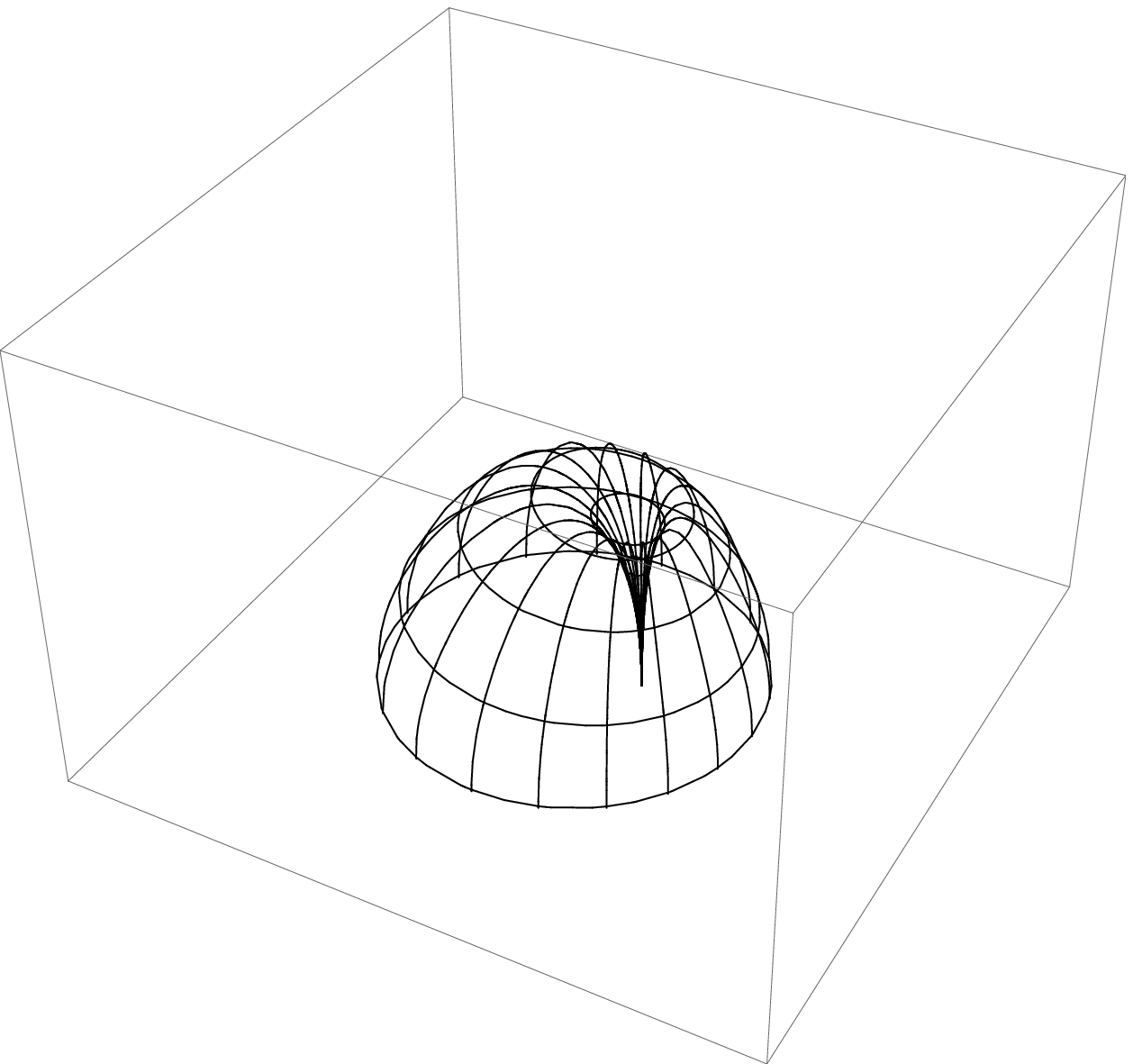}
\caption{The configuration for $(c_1, c_4) = (0.1,0.5)$
and $j=r=1$.}
\end{center}
\end{minipage}
\end{center}
\end{figure}
Then the Wilson loop is mapped to a circle 
which is parametrized by $\sigma$ inhomogeneously as
\begin{equation}
\vec x (\sigma) = \vec X'(\sigma) = 
{(r \cos \sigma + c_1 r^2,\,\, r \sin \sigma ,\,\, 0, \,\,  c_4 r^2)
\over 1 + 2 c_1 r \cos \sigma + {\vec c}^2 r^2 } \,.
\label{circle'}
\end{equation}
Because of the sigma dependence in $\vec \Theta(\sigma)$, 
this inhomogeneity cannot be removed from the Wilson loop
by reparametrization;
in other words,
the coupling to the scalar fields is changed from the original one.

The circle \eqref{circle'} is on 
the plane $\Sigma_{\vec c}$ which is specified by 
\begin{equation}
\Sigma_{\vec c}: \quad 
a {X^1}' +
b {X^4}' = 1\,, 
\quad
{X^3}' = 0, 
\quad
Y'=0
 \,,
\end{equation}
where,
\begin{equation}
a = 2 c_1, \quad 
b= {1 + \vec c^{\,2} r^2 - 2 (c_1)^2 r^2 \over c_4 r^2}\,.
\label{ab}
\end{equation}
The center $\vec X_W$ of the loop is now placed at 
\begin{equation}
\vec X_W
= 
\bigg(
{r^2 c_1 (-1 + r^2 {\vec c}^{\,2}) \over  (1 + r^2 {\vec c}^{\,2})^2 - (2 r
c_1)^2}, \,
0, \, 0, \,
{r^2 c_4 (1 + r^2 {\vec c}^{\,2}) \over  (1 + r^2 {\vec c}^{\,2})^2 - (2 r
c_1)^2}
\bigg)\,, 
\end{equation}
while the local operator is mapped to 
\begin{equation}
\vec x = 
\bigg(
{c_1 \over {\vec c}^{\,2} }\,, 0\,,0\,,{ c_4 \over {\vec c}^{\,2} }
\bigg)\,.
\end{equation}
In the following, we concentrate on the case
$(1 + r^2 {\vec c}^{\,2})^2 - (2 r c_1)^2 \neq 0$.
Each figure 2 and 3 is the string worldsheet for 
$(c_1, c_4) = (0.3, 0.2)$ and $(c_1,c_4) = (0.1,0.5)$
with $j=r=1$.
The lower plane in each figure is $\Sigma_{\vec c}$.
The vertical axis in the left figure is taken to be
$(a{X^1}' + b {X^4}'-1)/\sqrt{a^2+b^2}$ 
and that in the right figure is $Y'$. 
The configuration is specified by three parameters
as depicted in Figure 4.
The parameter $r'$ is the radius of the Wilson loop, 
and $\ell'$ is the distance between the point 
$\vec x$ and the surface $\Sigma_{\vec c}$. 
The third parameter $\rho'$ is the distance 
between the center of the Wilson loop 
and the local operator projected on $\Sigma_{\vec c}$.
These parameters are given by
\begin{align}
r' & = {r \over \sqrt{(1 + r^2 {\vec c}^{\,2})^2 - ( 2 r c_1)^2}}\,,  \\
\rho' & = {|c_1| \over {\vec c}^{\,2} \sqrt{(1 + r^2 {\vec c}^{\,2})^2 -
 ( 2 r c_1)^2}}\,, \\
\ell' & = {|c_4| \over {\vec c}^{\,2} 
\sqrt{(1 + r^2 {\vec c}^{\,2})^2 - ( 2 r c_1)^2}} \,.
\end{align}

\begin{figure}[htbp]
\begin{center}
\begin{minipage}{8cm}
\begin{center}
\includegraphics[width=5cm]{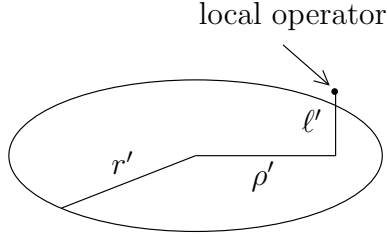}
\put(-70,79){local operator}
\put(-103,22){$r'$}
\put(-31,38){$\ell'$}
\put(-50,18){$\rho'$}
\caption{A generic configuration is 
specified by the parameters $r'$, $\rho'$ and $\ell'$.}
\end{center}
\end{minipage}
\end{center}
\end{figure}

For this solution, the evaluation of the string bulk action 
$S_{\rm bulk}$ is not changed from \eqref{Sbulk} because the isometry 
in AdS does not affect it.
On the other hand, the boundary terms do change.
The boundary term at $\tau_{\rm E}=\tau_-$ is changed as 
\begin{equation}
-
{\sqrt \lambda \over 2 \pi}
\int_0^{2 \pi} d \sigma 
{\dot Y' \over Y'} \bigg|_{\tau_-}
=
-
{\sqrt \lambda \over 2 \pi}
\int_0^{2 \pi} d \sigma 
\bigg[
{\dot Y \over Y}
-
{
\partial_{\tau_{\rm E}}
(1 + 2 \vec c \cdot \vec X + \vec c^{\,2} ( \vec X^2 + Y^2))
\over 
1 + 2 \vec c \cdot \vec X + \vec c^{\, 2} ( \vec X^2 + Y^2)
}\bigg]_{\tau_-}\,.
\end{equation}
The $\sigma$-integral exists because the S$^1$ 
symmetry of the original solution is lost, namely, 
$\vec c \cdot \vec X$ depends on $\sigma$.
However, the second term in the square bracket drops in the limit 
$\tau_- \to 0$ and the boundary term reduces to \eqref{b-term}.

The contribution from the 
vertex operator can be evaluated by using the relation
\begin{equation}
{Y' \over Y'^2 + (\vec X' - \vec x)^2}
=
{\vec c}^{\,2} Y\,.
\end{equation}
This relation holds for generic $\vec c$, in which 
case the position $\vec x$ of the local operator
is given by $\vec x = \vec c/ \vec c^{\,2}$.
Then by comparing it with the second expression in 
\eqref{EVAL-Vs1}, we understand 
that only the scaling factor is changed as follows:
\begin{equation}
\bigg( {r \over r^2 + \ell^2} \bigg)^J
\quad \to \quad 
\Big( r {\vec c}^{\,2} \Big)^J
=
\bigg(
{r' \over \sqrt{ 
(\rho'^2 + \ell'^2 - r'^2)^2 + 4 \ell'^2 r'^2}}
\bigg)^J\,.
\end{equation}
This is exactly the scaling behavior derived 
in \cite{Berenstein:1998ij}\cite{Alday:2011pf}.

\subsection{Second Solution}
\label{sec:Ano}

In \cite{Drukker:2006ga}, it was found that the unstable string solution
corresponding to the 1/4 BPS Wilson loop reproduces the contribution 
from the second saddle point for the Bessel function.
At the end of subsection \ref{sec:S-g}, we mentioned 
that there exists the second solution for the equation \eqref{EOMtE} 
which is given by
\begin{equation}
\sin \theta = 
{\sqrt {j^2 + 1} \over \cosh \sqrt{j^2 + 1}(-\tau_{\rm E} + \tau_0)}\,.
\label{theta2}
\end{equation}
In the case without angular momentum, i.e., $j=0$, 
the solution reduces to the one found in \cite{Drukker:2006ga}.
However, for $j \neq 0$, the ``size'' of the string becomes 
greater than the radius of S$^5$ in the range
$
\tau_0 - \xi/\sqrt{j^2+1} 
< \tau_{\rm E} < 
\tau_0 + \xi/\sqrt{j^2+1} 
$\,.
Hence we may expect that although it satisfies 
the required boundary conditions at $\tau_{\rm E}=0$
and $\tau_{\rm E} \to \infty$, it does not 
contribute to the semi-classical analysis of the string propagation.
On the other hand, in subsection \ref{sec:Sad},
we also found that the second saddle point 
which is expected to correspond to the solution \eqref{theta2}
is not on the steepest descent path we take.
If the string theory could reproduce the exact Bessel function
by the disk amplitude, it necessarily reproduce not only the leading 
contribution but also the whole structure of the integrand 
including the wrong saddle point. 
In this sense, the agreement of the behavior we found here seems to be fine.
Now let us check that the saddle point value of the 
integrand of the modified Bessel function
for this second saddle point is 
reproduced from the solution \eqref{theta2}.

For $\chi_2$, we have\footnote{
We set $\chi_2(0)=0$ for simplicity.
}
\begin{equation}
{\rm e}^{\pm i \chi_2} =  {\rm e}^{\pm j \tau_{\rm E}}
\bigg(
{
\sinh(\sqrt{j^2 + 1} (-\tau_{\rm E} + \tau_0) - \xi )
\over 
\sinh(\sqrt{j^2+1} (-\tau_{\rm E} + \tau_0) + \xi)
}
{
\sinh(\sqrt{j^2 + 1}\tau_0 + \xi)
\over 
\sinh(\sqrt{j^2 + 1}\tau_0 - \xi)
}
\bigg)^{\mp {1 \over 2}}\,,
\end{equation}
which has branch cut between two roots for $\sin \theta = 1$.
However, in terms of the embedding coordinates, 
${\cal X}_3(\tau_{\rm E}) \pm i {\cal X}_4(\tau_{\rm E})
= \cos \theta(\tau_{\rm E}) {\rm e}^{\pm i \chi_2(\tau_{\rm E})}$,
the branch cut coming from the factor ${\rm e}^{\pm i \chi_2}$
is canceled by the branch cut from 
the factor $\cos \theta$
\begin{equation}
\cos \theta {\rm e}^{\pm i \chi_2}
=
{\rm e}^{\pm j \tau_{\rm E}}
{
\sinh (\sqrt{j^2 + 1 }(-\tau_{\rm E} + \tau_0) \pm \xi)
\over 
\cosh \sqrt{j^2 + 1}(- \tau_{\rm E} + \tau_0)
}
\sqrt{
\sinh (\sqrt{j^2 + 1}\tau_0 \mp \xi)
\over 
\sinh (\sqrt{j^2 + 1}\tau_0 \pm \xi)
}\,.
\end{equation}

The evaluation of the bulk action is changed as follows
\begin{equation}
S_{\rm bulk} =
\sqrt \lambda \int_{\tau_-}^\infty d\tau_{\rm E}
(\sinh^2 \rho + \sin^2 \theta) =
\sqrt \lambda 
\bigg[{1 \over \tau_-} + \sqrt{j^2 + \cos^2 \theta_0} + \cdots
\bigg]\,.
\end{equation}
The S$^5$ part of the boundary term is also changed 
\begin{align}
J \log \cos \theta {\rm e}^{-i \chi_2} 
\bigg|_{\tau_+}
=
J
\bigg[
 - j \tau_+ - \log (\sqrt{j^2 + 1} - j)
- \log 
\Big(
\sqrt{ j'^2 + 1} - j'
\Big)
+
\pi i + \cdots
\bigg]
\end{align}
By adding all the contribution, we obtain
\begin{equation}
{\rm e}^{-S_{\rm total}}
\propto
(-1)^J
\exp \sqrt {\lambda'} 
\bigg[
- \sqrt{j'^2 + 1} - j' \log \Big( \sqrt{j'^2 + 1} - j' \Big)
\bigg]\,.
\end{equation}
It is exactly the saddle point value 
found in subsection \ref{sec:Sad}.
Since the AdS part of the vertex operator is not changed,
the scaling behavior found in subsection \ref{sec:Eva}
or subsection \ref{sec:Gen} is not affected.

\section{Summary and Discussion}
\label{sec:S-D}

We studied the holographic description
for the correlation function of the 
1/4 BPS Wilson loop and 1/2 BPS local operator
in the presence of the large R-charge.
First we constructed a rotating string solution which is 
extended in S$^5$ as well as AdS$_5$.
We checked that the solution preserves the 
1/8 of the supersymmetry as expected from 
the gauge theory computation.
This suggests that the correspondence may hold 
even for the finite distance between the 
Wilson loop and the local operator.
This means that the contribution from the descendant
operators in the OPE of the Wilson loop is
reproduced from the string computation.
By evaluating the total string action including 
boundary contributions, we found that the semi-classical
string amplitude reproduces the correct saddle point
value of the Bessel function.
The resulting expression is given by the same function
as \cite{Zarembo:2002ph} with the replacement 
\begin{equation}
\lambda' = \cos^2 \theta_0 \lambda\,, 
\qquad
j' = {j \over \cos \theta_0}\,. \notag 
\end{equation}
The first equation is expected 
from the work \cite{Drukker:2006ga}.
The correct scaling behavior is also reproduced.
Next we constructed string solutions for 
generic configuration of Wilson loop and 
local operator by using isometry of Poincar\'e AdS.
Because of the $\sigma$ dependence in $\vec \Theta (\sigma)$,
the Wilson loop operator is changed from the original one.
We have shown that the semi-classical analysis in the gravity side 
reproduces the scaling behavior, and hence
the contribution from the descendants, 
discussed in \cite{Berenstein:1998ij}\cite{Alday:2011pf}.

We also addressed the saddle point which is 
not on the steepest descent path in the gauge theory side.
In the string theory side we found the string saddle point
for which $\sin \theta$ becomes greater than 1, 
and hence it is not in the original integral domain.
In the case $J = 0$, the solution
reduces to the unstable solution discussed in \cite{Drukker:2006ga}.
This is quite natural since the corresponding saddle point 
of the Bessel function comes on the steepest descent path 
in the limit $j' \to 0$.
By evaluating the saddle point value both in the string theory
and the gauge theory, we found exact agreement for this 
saddle point as well.
This property would be a necessary condition 
for the correspondence to hold beyond the large $\lambda$ limit. \\

\noindent {\bf Note:}\quad
While we were preparing the manuscript, 
the paper \cite{Giombi:2012ep} appeared on arXiv,
in which correlation functions including
Wilson loop and local operator on S$^2$
is investigated, and the large $J$ 
behavior \eqref{e-S} is also derived 
in that context.

\section*{Acknowledgment}
We would like to thank 
Shinichi Deguchi, Koji Hashimoto, Tsunehide Kuroki, 
Shigefumi Naka, Takeshi Nihei, Satoshi Okuda, 
Satoshi Yamaguchi, for valuable discussions, 
suggestions and comments.

\end{document}